\documentclass{tcvg}               

\usepackage{mathptmx}
\usepackage{graphicx}
\newtheorem{algorithm}{Algorithm}
\usepackage{subfigure}
\usepackage{epsfig}
\usepackage{color}
\definecolor{dividerColor}{rgb}{0.6,0.6,0.6}

\onlineid{182}

\acmcategory{Research}
\acmformat{print}

\title{A parent-centered radial layout algorithm for interactive graph visualization and animation}

\author{Andrew Pavlo\thanks{e-mail:pavlo@cs.wisc.edu}\\ %
        \scriptsize University of Wisconsin-Madison %
\and Christopher Homan\thanks{e-mail:cmh@cs.rit.edu}\\ %
     \scriptsize Rochester Institute of Technology %
\and Jonathan Schull\thanks{e-mail:jis@it.rit.edu}\\ %
     \scriptsize Rochester Institute of Technology} %


\begin{document}

\maketitle


\begin{abstract}

\copyrightspace
 
We have developed (1) a graph visualization system that
allows users to explore graphs by viewing them as a succession of
spanning trees selected interactively, (2) a radial graph
layout algorithm, and (3) an animation algorithm that generates
meaningful visualizations and smooth transitions between graphs while
minimizing edge crossings during transitions and in static layouts.

Our system is similar to the radial layout system of Yee et
al.~\cite{yee01}, but differs primarily in that each node is
positioned on a coordinate system centered on its own parent rather
than on a single coordinate system for all nodes. Our system is thus
easy to define recursively and lends itself to parallelization. It
also guarantees that layouts have many nice properties, such as: it 
guarantees certain edges never cross during an animation.

We compared the layouts and transitions
produced by our algorithms to those produced by Yee et al.
Results from several experiments indicate that our system produces fewer edge crossings during
transitions between graph drawings, and that the transitions more
often involve changes in local scaling rather than structure.

These findings suggest the system has promise as an interactive graph
exploration tool in a variety of settings.

\end{abstract}

\begin{CRcatlist}
  \CRcat{I.3.3}{Computer Graphics}{Picture/Image Generation}{Viewing algorithms};
  \CRcat{H.5.0}{Information Interfaces and Presentation}{General}{}
\end{CRcatlist}

\keywordlist

\section{Introduction}
Visualization can help make graph structures
comprehensible~\cite{herman00,larkin87,tufte83}. However, edge crossings can
challenge human perception of relationships between
nodes~\cite{huang05,purchase98,ware02}, yet graphs often come to us as
tangled webs that cannot be depicted without crossings in a two-dimensional viewing plane.

Because trees \emph{can} be laid out on a plane without edge
crossings, a common approach is to base graph visualizations on
spanning trees extracted from
graphs~\cite{jankunkelly03,munzner95,noel02,yee01}. Although the
resulting drawings may discard some potentially significant edge
information, a clearer mental picture of the full graph may
nonetheless result if users can easily and intuitively explore
multiple layouts based on different spanning trees.

\begin{figure}[t]
	\centering
	\small
	\subfigure[]{
		\label{fig:screenshot1.1}
		\includegraphics[width=.22\textwidth]{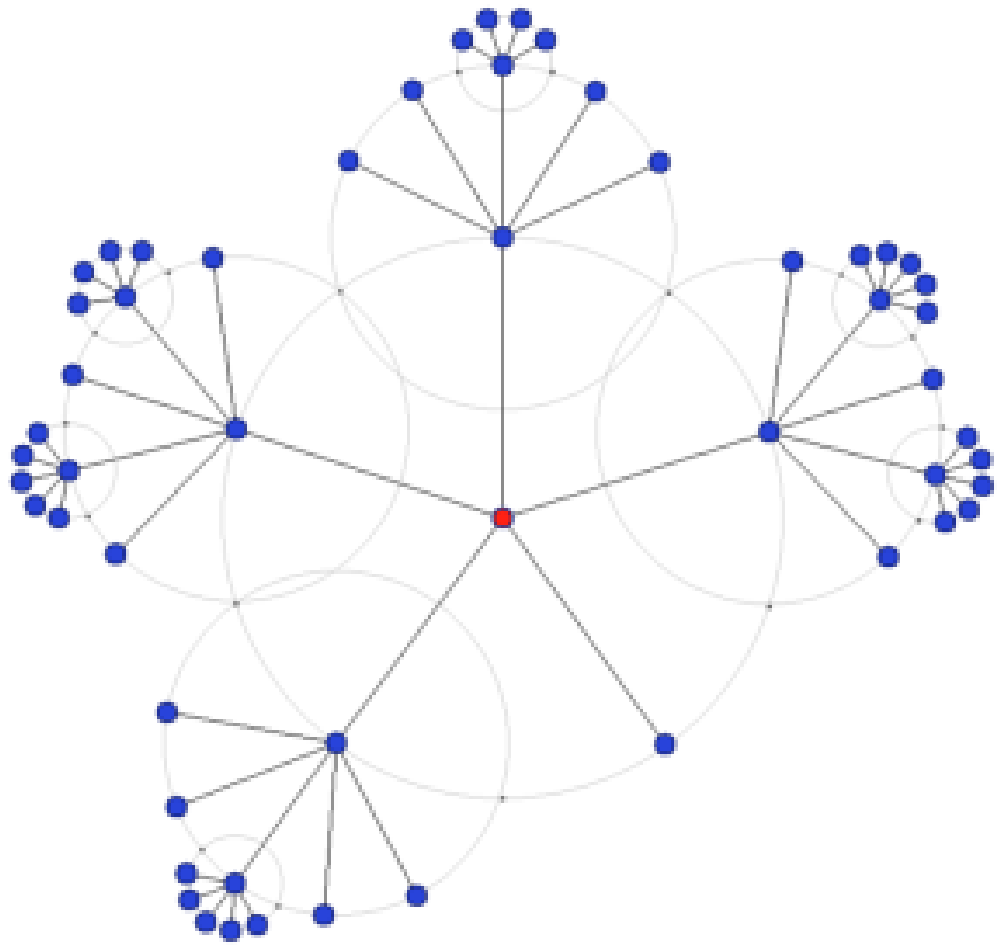}
	}
	\subfigure[]{
		\label{fig:screenshot1.2}
		\includegraphics[width=.22\textwidth]{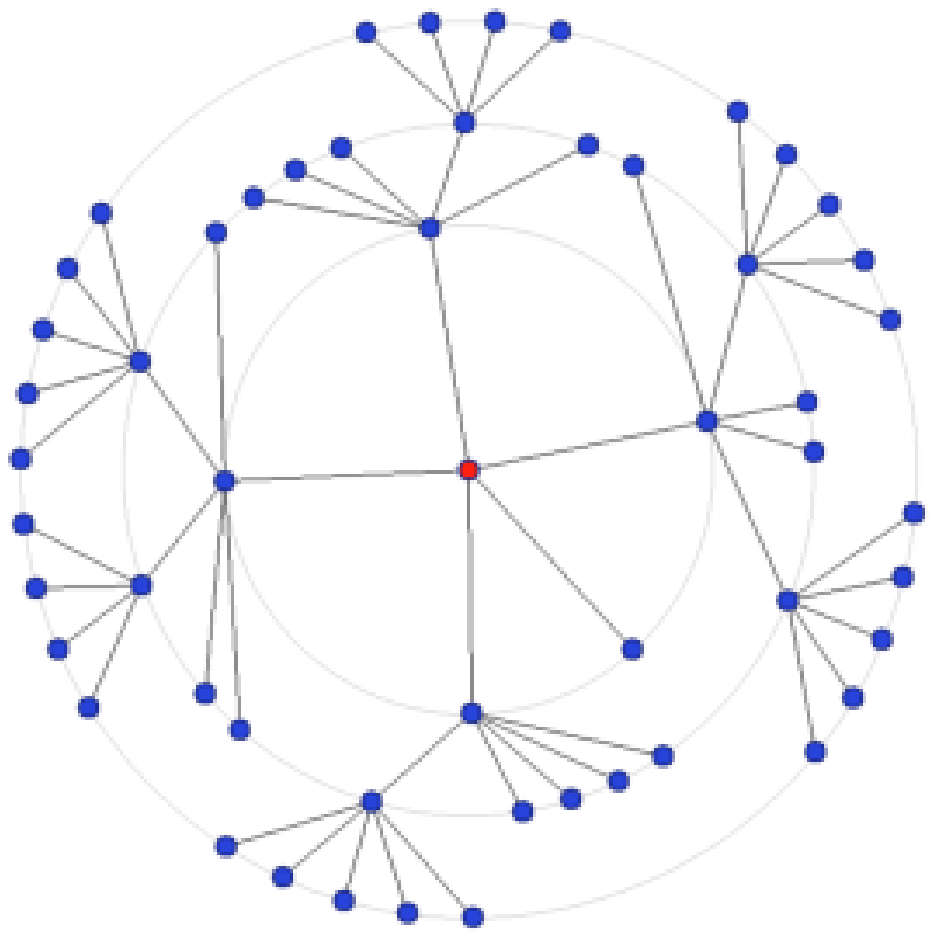}
	}
	\caption{Two drawings for the same tree: Figure~\ref{fig:screenshot1.1} is drawn by our layout algorithm, and Figure~\ref{fig:screenshot1.2} is drawn by Yee et al.'s layout algorithm for Gnutellavision~\cite{yee01}.}
	\label{fig:screenshot1}
\end{figure}

Yee et al.~\cite{yee01} describe a tool that draws radial tree
layouts~\cite{eades92,melancon98,teoh02,wills99} of breadth-first
spanning trees, given a graph and a node selected to be the root (see
Figure~\ref{fig:screenshot1.2}). A user may then select a new root
node and the system transitions smoothly to a new layout based on
the new root node. This transition is animated by a succession of
linear interpolations of the polar coordinates of positions of each node
in the old and new layouts. Thus, a user can
interactively explore a graph that would otherwise be too complex to
visualize or comprehend as a single, static drawing.

In drawings generated by Yee et al.'s radial layout method, 
successive generations of nodes lie on concentric circles centered on the
root. Such layouts guarantee that all nodes of a given
generation are equidistant from the root and lend themselves to a smooth animation process. However, symmetries in the tree can be obscured because distantly related nodes may be
positioned close to each other in the final layout just because they belong to the same
generation. And, during animations, numerous edge crossings may occur.

\begin{figure*}[t]
	\centering
	\small
	\subfigure[]{
		\label{fig:layout1.1}
		\includegraphics[width=.24\textwidth]{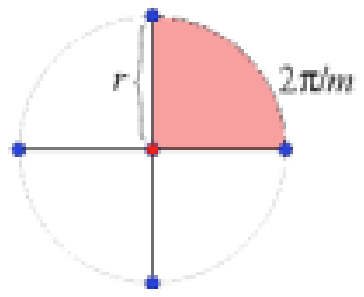}
	}
	\subfigure[]{
		\label{fig:layout1.2}
		\includegraphics[width=.24\textwidth]{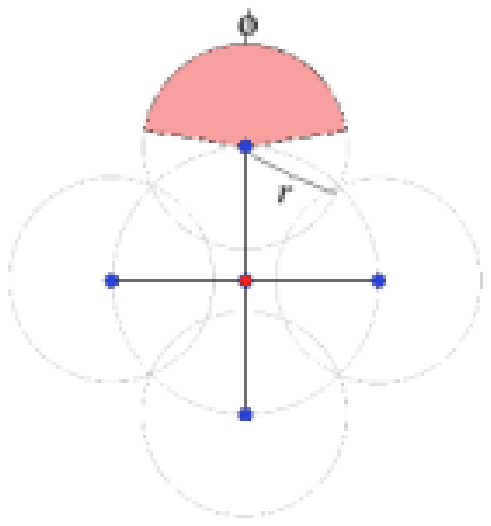}
	}
	\subfigure[]{
		\label{fig:layout1.3}
		\includegraphics[width=.24\textwidth]{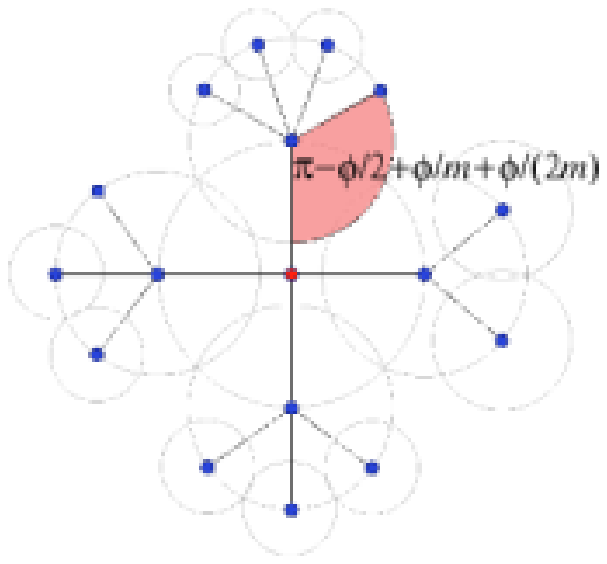}
	}
	\subfigure[]{
		\label{fig:layout1.4}
		\includegraphics[width=.24\textwidth]{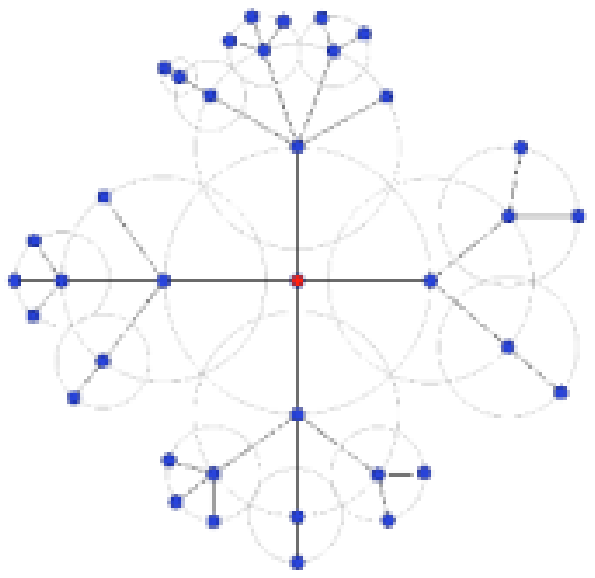}
	}
	\caption{A visual overview of how our algorithm constructs a new layout for a graph. In Figure~\ref{fig:layout1.1}, the root is first placed at the center of the drawing surrounded by its children. Next, the root's children are allocated containment arcs in Figure~\ref{fig:layout1.2} where their descendants are positioned within Figure~\ref{fig:layout1.3}. The final static layout is shown in Figure~\ref{fig:layout1.4}.}
	\label{fig:layout1}
\end{figure*}

Our approach is similar to Yee et al.'s
in that it bases its drawings on breadth-first rooted spanning trees
extracted from graphs, allows users to interactively change views of
each graph by selecting a new root, and smoothly transitions between
successive layouts by moving nodes along radial paths.   
However, unlike Yee et al., we place every subtree in the graph in a ``parent-centric'' circle surrounding its own subroot, instead of positioning each node on a ``generation circle'' centered on the root. This approach lends itself naturally to recursion, and naturally reflects the self-similar structure of recursively branching trees.   
Moreover, it guarantees that during the animation process the edges between 
parent and child never cross, a property Yee et al.'s algorithm does not provide.
From an algorithmic perspective, each node's position depends only on its parent and siblings, not on its entire generation.  Because the dependencies in our layout are therefore very local, drawings and animations in our system are potentially computable in a single, 
parallelizable, traversal of the tree.

In broad strokes, our layout algorithm works as follows. First, we place the root in the center of the display with its children evenly distributed along a \emph{containment circle} centered on the root.  Second, we draw circles around the root's children and evenly distribute \emph{their} children along \emph{containment arcs} that ensure that neither siblings nor cousins overlap. Then the second process just proceeds recursively, so that successively distant descendants of the root are positioned on successively smaller containment arcs (Figure~\ref{fig:layout1}). 

Our layouts have several aesthetic virtues: They have a flowerlike, self-similar structure that differs from the ``bulls-eye'' appearance of Yee et al.'s layouts (compare Figure~\ref{fig:screenshot1.1} and Figure~\ref{fig:screenshot1.2}). And even though the distance between root and nonroot nodes is less directly represented than in Yee et al.'s system, there are powerful visual cues to compensate: Within a lineage, edge lengths decrease monotonically with distance from root, and all siblings within a family are arrayed along visually salient arcs equidistant from their common parent. Also, our out-facing containment \emph{arcs} ensure that the correlation between graph distance and Euclidean distance from the root is more  reliable than in  parent-centered approaches based on containment \emph{circles}~\cite{melancon98,quang02,teoh02}. With regard to animating transitions, our algorithm ensures that sibling edges never cross when a new focal node is selected, and whenever the graph to be drawn is itself a tree. 

\section{Data Model and Algorithms}
\label{sec:method}
We assume that all graphs are connected, and regard any drawing of a spanning tree of a graph as a drawing of the graph. Since all edges are to be drawn as
straight lines, we need only describe the mapping of nodes to points in the drawing plane (we use ``node'' to refer to
both a vertex in a graph and to the location of the node on the drawing plane.)  It is perhaps easiest to explain our algorithms in terms of
a particular data model that completely describes a drawing in this
restricted sense. 

Rather than represent the position of all nodes
of some graph in terms of a single polar coordinate system centered at the 
origin of
the drawing plane that all nodes share, we only use the standard drawing-plane's 
origin to represent the root node and its children. We represent every
other node position in terms of polar coordinates \emph{centered at
the node's parent}~\cite{kreuseler00}.

\subsection{Parent-centered data model}
We now formally define this concept.  Given a tree $T$ and a drawing $D$
of $T$, we recursively
define a \emph{parent-centered model} of $(D,T)$ as follows. For any node $v$ of $T$,
the polar coordinates of $v$ are given in the coordinate system
\begin{description}
  \item[(basis 1, i.e., if $v$ is the root of $T$:)] sharing the origin
with the drawing plane and zero degrees with the positive direction of
the drawing plane's $x$-axis,

\item[(basis 2, i.e., if $v$ is a child of the root of $T$:)] having the root of $T$ as the origin and zero degrees as the ray from the root having the same direction as the positive direction of the drawing plane's $x$-axis, or 
\item[(recursion, i.e., otherwise:)] having $v$'s parent in $T$ as the
origin and the ray from $v$'s parent intersecting $v$'s grandparent as zero degrees.
\end{description}
Thus nodes having the same parent share the same coordinate system and 
nodes having different parents have different coordinate systems.

This data model applies to the static and dynamic layout algorithms described below. Note that we can (and do) represent any straight-line graph drawing
this way, not just those produced by Algorithm~\ref{alg:static} below.

\subsection{Static layout algorithm}
We define our static layout algorithm recursively as follows (see also Figure~\ref{fig:layout1}).  
When we say that a nonroot node lies on a \emph{containment circle}, we are referring to the circle centered at the node's parent that intersects the node. Note that if two siblings are the same distance from their parent (this is a property of the drawings Algorithm~\ref{alg:static} produces) then they share the same containment circle.

\begin{algorithm}\label{alg:static}
Given a spanning tree $T$, for each node $v$ of $T$ let the coordinates of the root node be $(0,0)$ and for each nonleaf node $v$ let $v_1,\ldots,v_m$ be $v$'s children. For each $i \in \{1,\ldots,m\}$ let the coordinates of $v_i$ (in the parent-centered model) be 

\begin{description}
	\item[(basis, i.e., if $v$ is the root:)] $(2\pi i/m, r)$, where $r$ is some user-defined value $> 0$,

        \item[(recursion, i.e., otherwise:)] $(\pi - \phi/2 +  \phi i/m + \phi/(2m), r)$, where $\phi$ is some user-defined value $> 0$ and $r$ is 
          \begin{itemize}
            \item half of $v$'s magnitude, if $v$ has no siblings, otherwise
            \item the radius of the circle centered at $v$ that intersects the midway point between $v$ and $v$'s nearest sibling(s) on their shared containment circle. 
              \end{itemize}
\end{description}
\end{algorithm}
Note that the value of $r$ for any nonroot node depends only on the node's parent, so as claimed above all sibling nodes share the same value for $r$. This means they all lie on the same containment circle, which we call the containment circle \emph{of} the parent node.

\subsection{Animation algorithm}
\label{sec:method_animation}
\begin{figure*}[t]
	\centering
	\small
	\subfigure[]{
		\label{fig:screenshot4.1}
		\includegraphics[width=.185\textwidth]{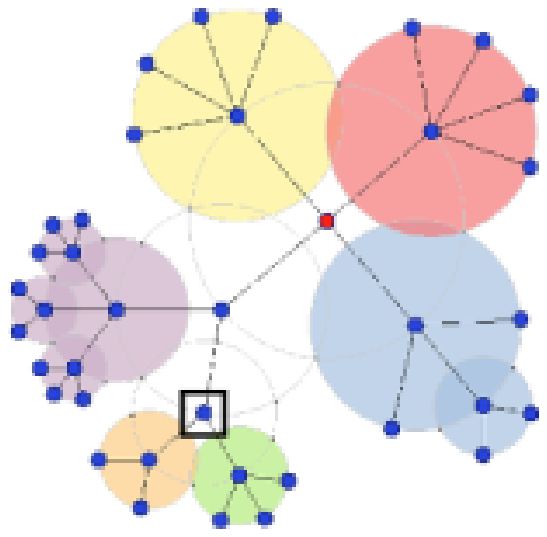}
	}
	\color{dividerColor}
	\line(0,1){90}
	\normalcolor
	\subfigure[]{
		\label{fig:screenshot4.2}
		\includegraphics[width=.185\textwidth]{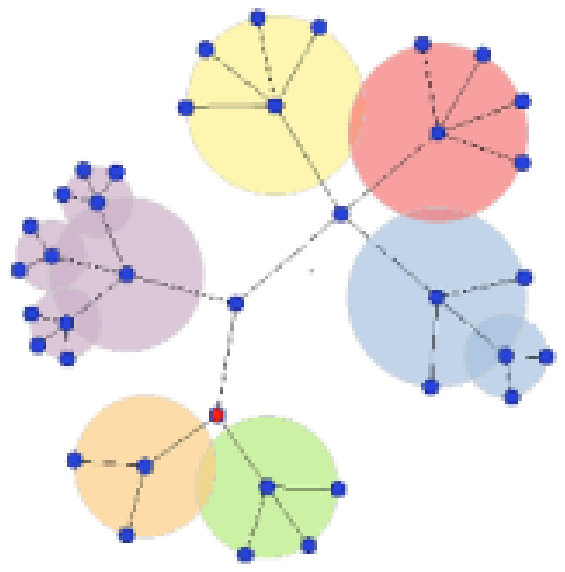}
	}
	\color{dividerColor}
	\line(0,1){90}
	\normalcolor
	\subfigure[]{
		\label{fig:screenshot4.3}
		\includegraphics[width=.185\textwidth]{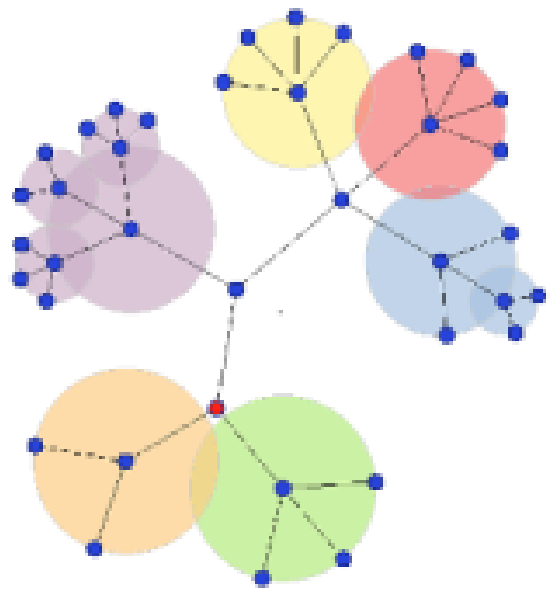}
	}
	\color{dividerColor}
	\line(0,1){90}
	\normalcolor
	\subfigure[]{
		\label{fig:screenshot4.4}
		\includegraphics[width=.185\textwidth]{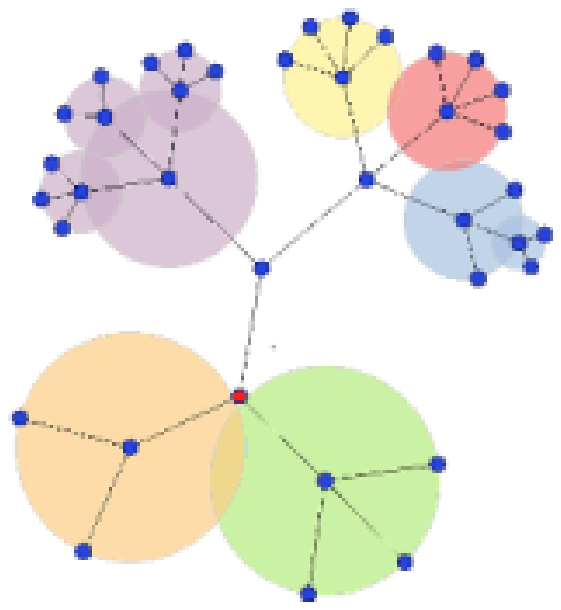}
	}
	\color{dividerColor}
	\line(0,1){90}
	\normalcolor
	\subfigure[]{
		\label{fig:screenshot4.5}
		\includegraphics[width=.185\textwidth]{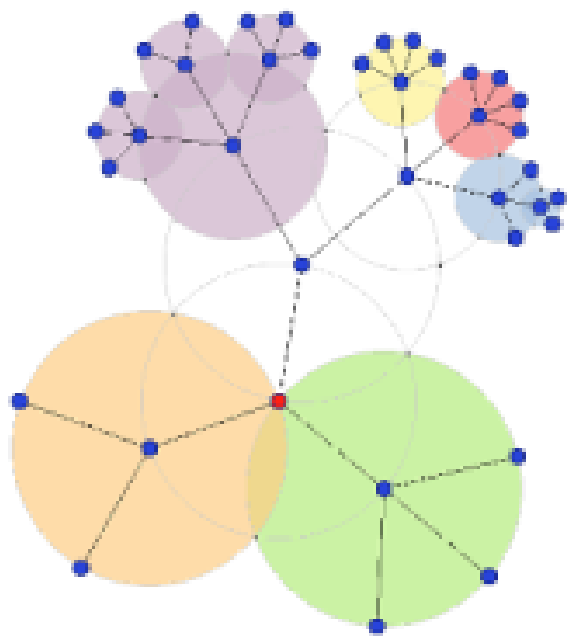}
	}
	\caption{Isomorphic tree transition -- Our visualization scheme transitions between two drawings of the same tree by scaling each parent node's containment circle with its children. In this example, the user selects a new root node from the initial drawing in Figure~\ref{fig:screenshot4.1}. The containment circles highlighted during the transition in Figures~\ref{fig:screenshot4.2} to \ref{fig:screenshot4.4} grow and shrink as the graph moves to the new drawing in Figure~\ref{fig:screenshot4.5}.}
	\label{fig:screenshot4}
\end{figure*}

\begin{figure*}[t]
	\centering
	\small
	\subfigure[]{
		\label{fig:screenshot3.1}
		\includegraphics[width=.185\textwidth]{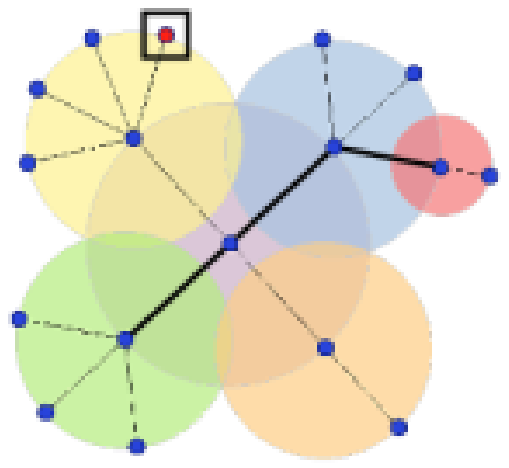}
	}
	\color{dividerColor}
	\line(0,1){90}
	\normalcolor
	\subfigure[]{
		\label{fig:screenshot3.2}
		\includegraphics[width=.185\textwidth]{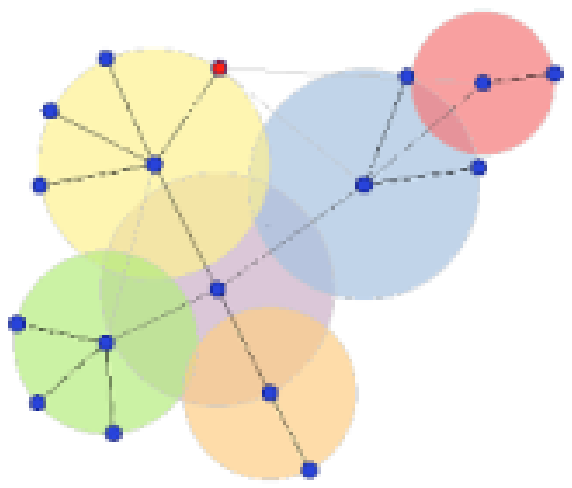}
	}
	\color{dividerColor}
	\line(0,1){90}
	\normalcolor
	\subfigure[]{
		\label{fig:screenshot3.3}
		\includegraphics[width=.185\textwidth]{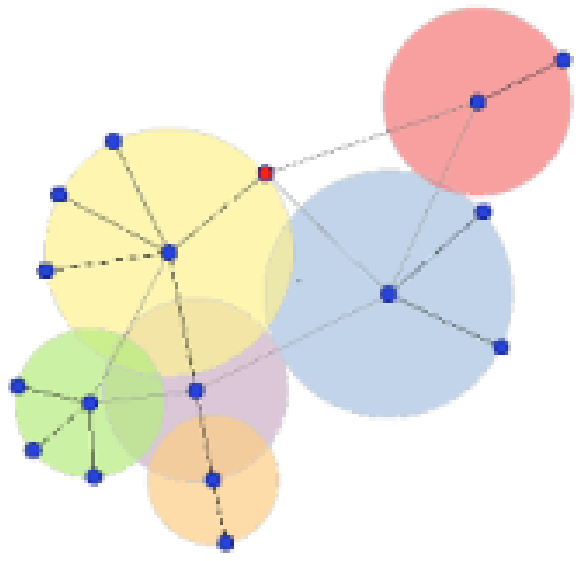}
	}
	\color{dividerColor}
	\line(0,1){90}
	\normalcolor
	\subfigure[]{
		\label{fig:screenshot3.4}
		\includegraphics[width=.185\textwidth]{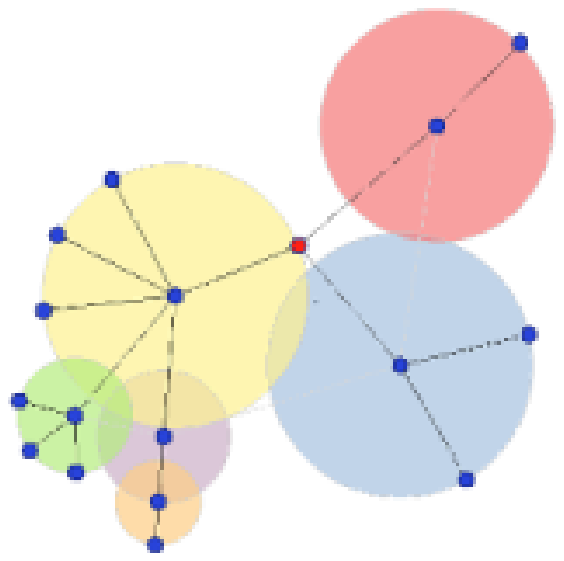}
	}
	\color{dividerColor}
	\line(0,1){90}
	\normalcolor
	\subfigure[]{
		\label{fig:screenshot3.5}
		\includegraphics[width=.185\textwidth]{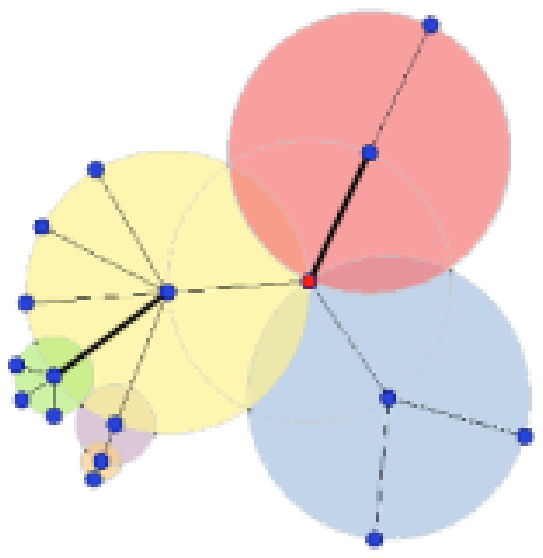}
	}
	\caption{Spanning-tree-to-spanning-tree transition -- The user's selection of a new node invokes a new spanning-tree-based drawing. Edges that will fade out are highlighted in Figure~\ref{fig:screenshot3.1}; newly introduced edges are highlighted in Figure~\ref{fig:screenshot3.5};}
	\label{fig:screenshot3}
\end{figure*}

Our static layout algorithm leads to a simple and intuitive
algorithm for animating transitions from one layout to another by interpolating between the parent-centered models of $D_{old}$ and a drawing produced by Algorithm~\ref{alg:static} of a spanning tree of $G$ rooted at $v$, for
any graph $G$, drawing $D_{old}$ and node $v$ of $G$ (see Figures~\ref{fig:screenshot3}
and~\ref{fig:screenshot4}).

\begin{algorithm}\label{alg:anim}
\begin{enumerate}
\item Compute a breadth-first spanning tree $T$ of $G$ rooted at $v$.
\item Let $D_{new}$ be a drawing produced by running
Algorithm~\ref{alg:static} on $G$ and $T$.
\item Let $M_{old}$ be a parent-centered model of $(D_{old}, T)$ and
$M_{new}$ be a parent-centered model of $(D_{new}, T)$
\item For each $t$ in an increasing sequence $0, t_1,\ldots, t_p,1$,
output a polar drawing $D_t$ such that the model of $(D_t,T)$ is
described recursively as follows.
\begin{description} \label{alg:main}
\item[(basis, i.e., if $v$ is the root of $T_{new}$):] $(\theta,
(1-t)r)$, where $(\theta, r)$ are the polar coordinates of $v$ in
model of $M_{old}$.
\item[(recursion, i.e., otherwise:)] $(t\theta_{new} +
(1-t)\theta_{old}, tr_{new} + (1-t)r_{old})$ otherwise, where, for $x
\in \{old,new\}$, $(\theta_x,r_x)$ are the coordinates of $v$ in
$M_x$.
\end{description}
\end{enumerate}
\end{algorithm}

Thus, the new root node moves in a straight line to the center of the
new drawing, and each nonroot node moves via a finite approximation of
a smooth interpolation between its parent-centered polar coordinates
in the new and old drawings.  In the resulting animation,
newly-central families expand and fan out as they move toward the
center, while newly-peripheral families shrink as they arc toward the
periphery.  Neighboring family circles are guaranteed not to
interpenetrate. Note that any model $M_t = (D_t,T)$ will not generally look
like a model generated by Algorithm~\ref{alg:static} since, for
instance, the root of $T$ may not in $M_t$ lie on the origin of the drawing
plane.

Our algorithm is built into a system that first
displays a forced-directed layout~\cite{eades84} $D_{old}$ of a given graph $G$. A user then
clicks on a node $v$ and the system runs Algorithm~\ref{alg:anim}, the
output of which, $D_{new}$, is set to $D_{old}$ the next time Algorithm~\ref{alg:anim} is called, which is the next time a user clicks on a node.
 Thus, $D_{old}$ is 
typically a drawing produced by Algorithm~\ref{alg:static}

(though it need not be).

There are different ways in which one can fix the times $t_1,\ldots, t_p$ 
when generating the intermediate drawings of an animation.
We adopted the slow-in, slow-out technique of Yee et al. in our implementation
so that the values of $t_1,\ldots, t_p$ are concentrated toward the boundary values $0$ and 
$1$.

\subsection{Properties}
Our algorithms have some noteworthy properties.
\begin{description}
\item[Aesthetics:] 
Our layout algorithm (1) ensures that all siblings are equally distant from their parent, (2) ensures that containment arcs of siblings and cousins do not overlap, and (3) produces layouts that provide clear indications (via edge-length and family shape) of closeness to the root.  

Our animation process also guarantees that certain edges never cross. For any graph $G$, any drawing $D_{old}$ and a node $v$ of $G$, there is a choice for $\phi$ such that for any time $t \in [0,1]$, the edges corresponding to the spanning tree upon which $D_t$ is based do not cross in $D_t$. This has a number of consequences, including:
\begin{enumerate}
\item If $G$ is a tree then the drawing $D_t$ has no edge crossings.
\item For any node $v$ of $G$, the edges between $v$ and its children do not cross.
\end{enumerate}
In the next section we describe experiments that test how well Algorithm~\ref{alg:anim} avoids edge crossings overall.
\item[Parallelizability:]
Note that all four steps of Algorithm~\ref{alg:anim} can be implemented as a single traversal of $T$ (i.e., during the breadth-first search that produces $T$). 
Algorithm~\ref{alg:anim} thus lends itself easily to parallelization, as a new process can be forked whenever a node of $T$ is traversed. 
\end{description}

\section{Experiments}
\label{sec:experiments}
Our experiments compare our algorithms' layouts and animations to those produced by Yee et al.'s algorithms. In each experimental trial, a random graph was generated, two distinct root nodes within that graph were randomly selected, and the graph was then operated upon by both algorithms as they effected transitions from a spanning tree rooted at the first node to a spanning tree rooted at the second. Each experiment comprised 710 trials per algorithm, in which ten random graphs of order 30--100 (inclusive) were generated using the Erd\"os-R\'enyi model~\cite{erdos60} with a 10\% probability of an edge connecting any two nodes. In our first two experiments, we counted edge crossings during transitions by examining all edges present during the transition (whether derived from the new spanning tree or the old); a single crossing was counted during a trial if two edges crossed at any time during the transition, even if the edges crossed and uncrossed multiple times. In our last experiment, we measure the lengths of edges for sets of sibling nodes to their common parent in static layouts produced by the algorithms.

\subsection{Isomorphic tree transitions}
\label{sec:experiment1}
Because trees are by definition planar, transitional edge crossings are potentially avoidable in the special case where selection of a new root node does not change a tree's edge set. In this experiment, we first extract a spanning tree from a graph rooted at a randomly selected root node and construct a new drawing. We then transition from this drawing to a second drawing of the same tree but with a different node selected as the focal point.

\begin{figure}
	\small
	\centering
	\includegraphics[width=.40\textwidth]{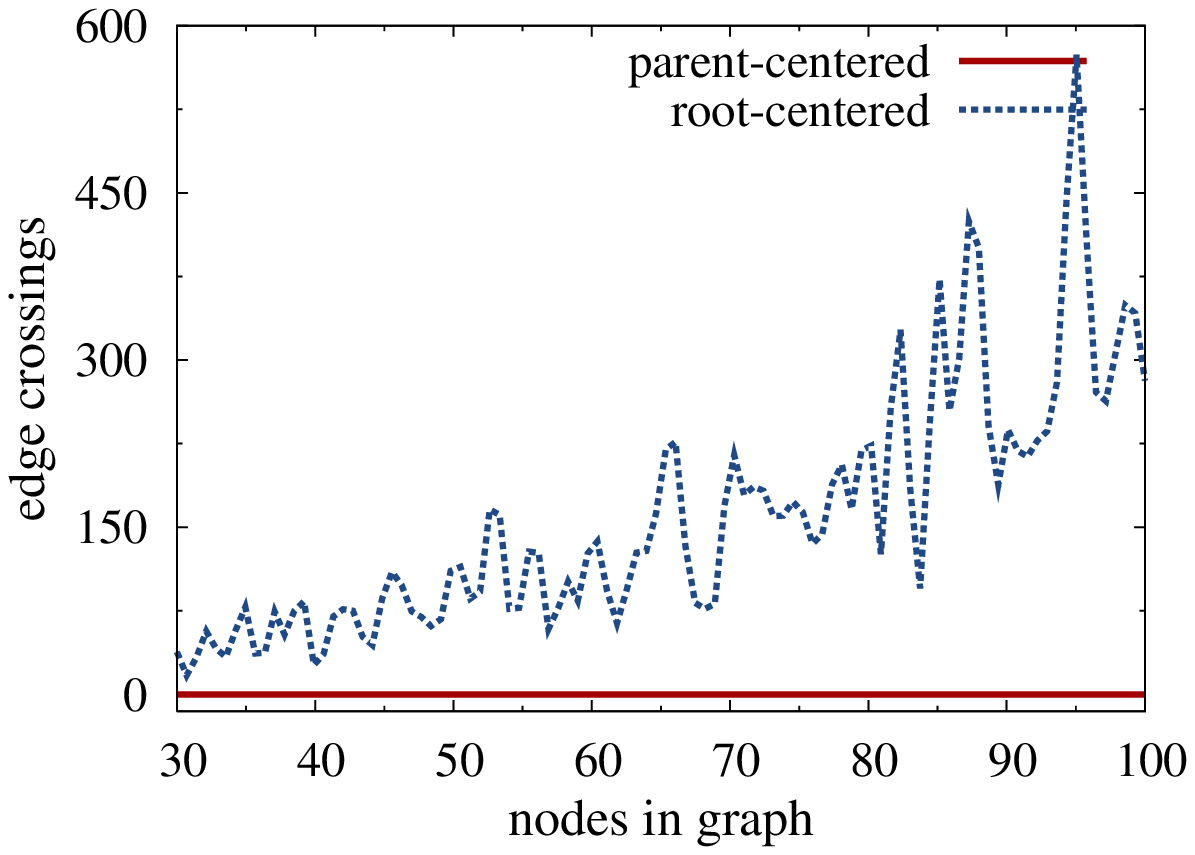}
	\caption{Isomorphic tree transitions \newline Our parent-centered visualization scheme produces no edge crossings when transitioning between drawings of the same tree, while Yee et al.'s root-centered system produces many.}
	\label{fig:experiment1}
\end{figure}

Figure~\ref{fig:experiment1} shows that our algorithms successfully produce zero crossings while Yee et al.'s algorithms produce many for this particular transition scenario. As illustrated in Figure~\ref{fig:screenshot4}, our approach avoids crossings because ``family circles'' simply expand or contract as they move without interpenetrating. In contrast, Yee et al.'s algorithms maintain visual continuity by preserving the direction of the edge from the new root node to its parent in the previous drawing. This can produce dramatically different drawings of the same tree, and result in crossings during the transitions.

This visual effect of our animations is similar to that of rigid-body animation methods~\cite{friedrich02.1,friedrich02.2} as the user can mentally group subgraphs as separate objects and follow the movements more easily \cite{nesbitt02}.

\subsection{Spanning-tree-to-spanning-tree transitions}
\label{sec:experiment2}
In the second experiment, we counted edge crossings during transitions between two different spanning-tree-based drawings of the same graph. We first create a spanning-tree-based drawing for a graph rooted at a randomly selected node. We then select a second node for a new drawing based on a different spanning tree extracted from the graph. Unlike in the previous experiment, the edge sets of the two drawings are not the same in this experiment.

Our evaluation distinguishes ``transient crossings involving fading-out edges'' from ``transient crossings involving final layout edges''. A crossing is \emph{transient but fading} if at least one of the edges fades from the viewing plane during the animation sequence. A \emph{transient and non-fading} crossing occurs when both edges are part of the final drawing.

\begin{figure}[t]
	\small
	\centering
	\subfigure[Transient fading-out crossings]{
		\label{fig:experiment2.1}
		\includegraphics[width=.40\textwidth]{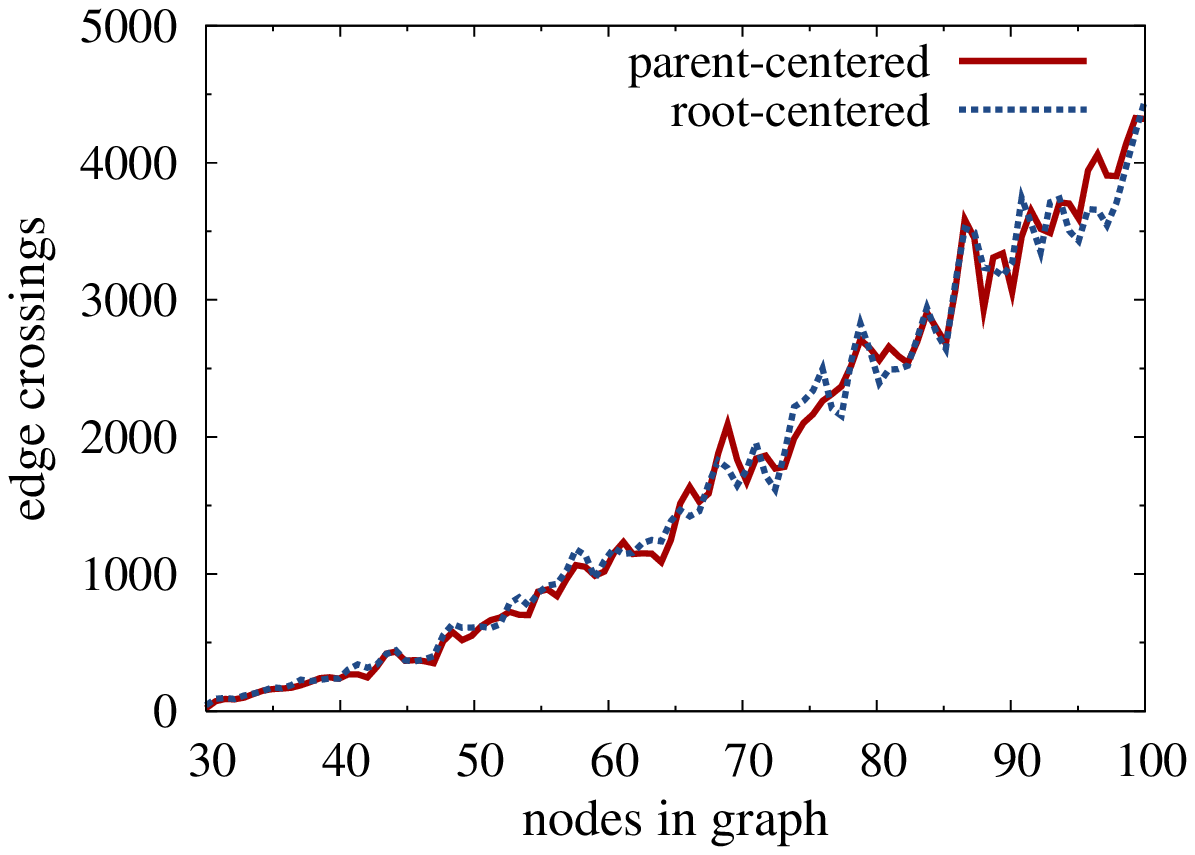}
	}
	\subfigure[Transient final layout crossings]{
		\label{fig:experiment2.2}
		\includegraphics[width=.40\textwidth]{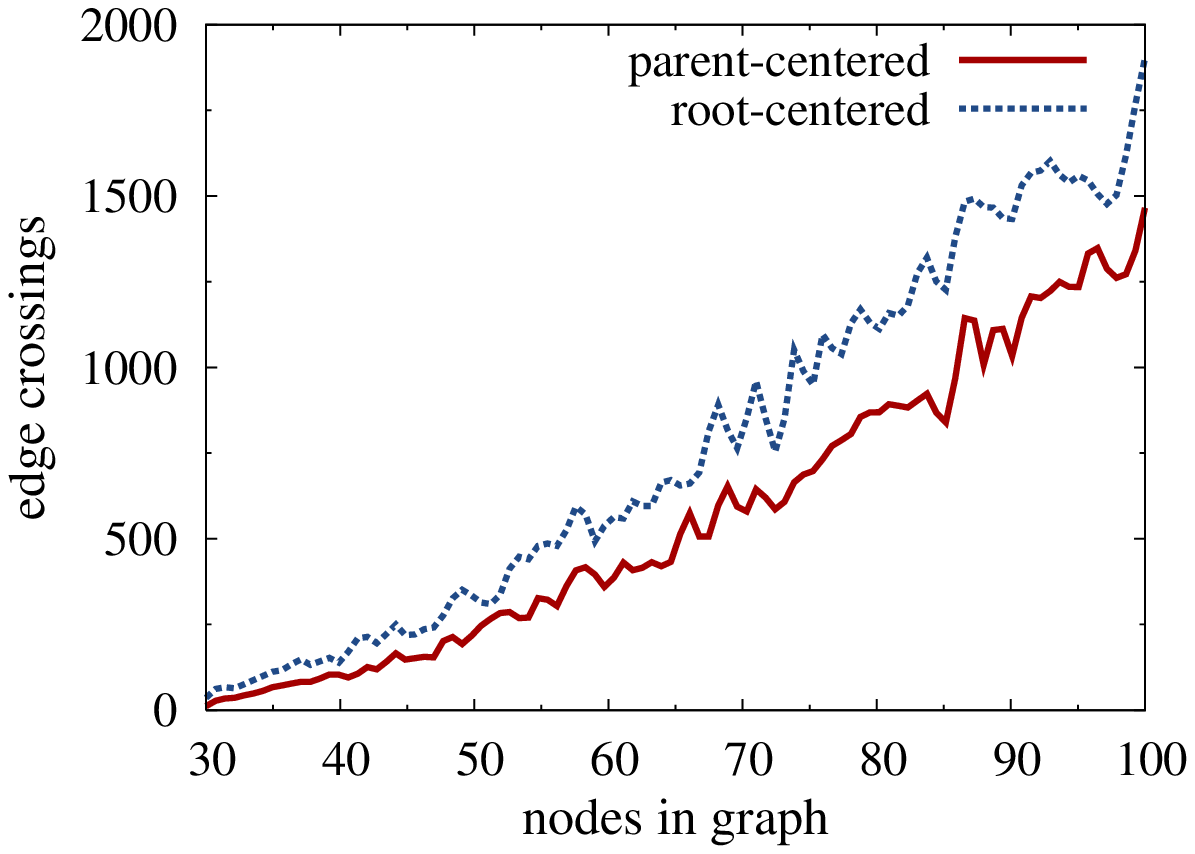}
	}
	\caption{Spanning-tree-to-spanning-tree transitions \newline The results in Figure~\ref{fig:experiment2.1} shows that both visualization schemes produce similar amounts of edge crossings during transitions between two different spanning-tree-based drawings. The results in Figure~\ref{fig:experiment2.1}, however, clearly show that our parent-centered algorithms produced fewer crossings than Yee et al.'s root-centered algorithms.}
	\label{fig:experiment2}
\end{figure}

As shown in Figure~\ref{fig:experiment2.1}, the two visualization schemes produce a comparable number of transient but fading crossings. But Figure~\ref{fig:experiment2.2} shows that our algorithms produce fewer transient and non-fading crossings than Yee et al.'s, and that this difference grows with graph order.

\subsection{Spanning tree sibling edge lengths}
\label{sec:experiment3}
Since our approach positions nodes on containment arcs around their parent whereas Yee et al. positions nodes on concentric circles around the root node, the two systems produce different patterns of regularities. In this experiment we quantify those regularities.

\begin{figure}
	\small
	\centering
	\includegraphics[width=.40\textwidth]{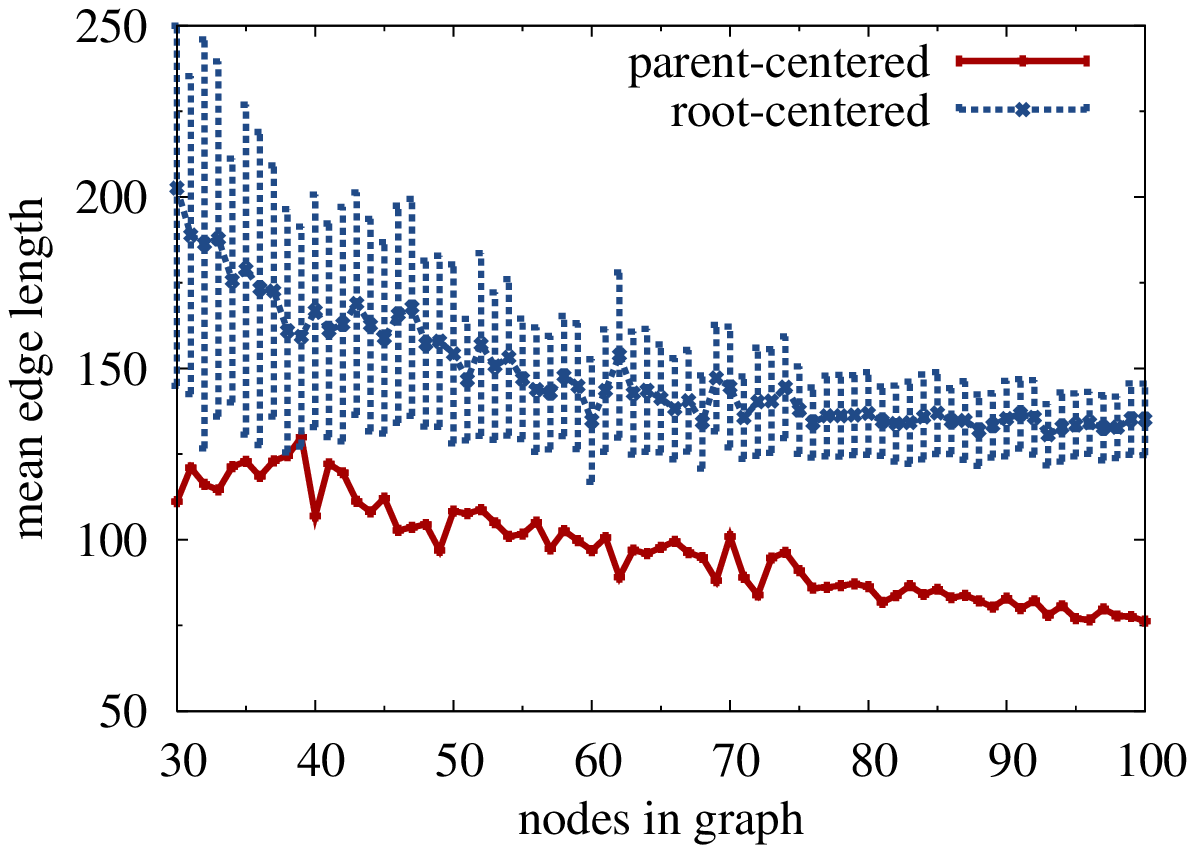}
	\caption{Spanning tree sibling edge lengths \newline Edge lengths from siblings to their common parent tend to decrease as graph size increases.  In our algorithms, siblings are positioned equidistant to their common parent.   In Yee et al.'s algorithms, edge lengths from siblings to common parent vary, as shown by standard deviations.}
	
	\label{fig:experiment3}
\end{figure}

Figure~\ref{fig:experiment3} shows that as the generational distance increases from the root to nodes at a given depth in the tree, our system produces no variance among siblings in within-family distance from node to parent, whereas Yee et al.'s system produces substantial variance. 

Conversely, in our system the distance from the root to nodes of a given generation can vary, whereas in Yee et al.'s system it does not. The variance arises in our case because our algorithm adjusts containment arcs to help prevent neighboring family circles from overlapping. Although this reduces the reliability of the length of edges as an indicator of distance from the root, the self-similar geometric pattern of family subsystems produces another cue that may well be more salient~\cite{montello03}.

\section{Discussion and Future Work}
\label{sec:discussion}
Behavioral tests will be needed to determine whether these alternative layout and transition algorithms are psychologically significant. But our statistical experiments indicate that the drawings and animated transitions generated by our algorithms conform to many established aesthetics for graph drawings~\cite{graphdrawing99,purchase98}.

Taken in the context of the prior research on graph drawing aesthetics, these results suggest that our system should reduce a user's mental effort and increase a user's capacity to make reliable judgments and develop useful intuitions about complicated graph structures~\cite{huang05,purchase98,ware02}. Our research thus lays the groundwork for future study of the layout and animation algorithms, of the psychological significance of our metrics, and of the functional validity of the graph aesthetics themselves.

With regard to our algorithms, two areas are particularly ripe for further study. First, the drawings produced by both ours and Yee et al.'s graph layout algorithm are not guaranteed to be planar; in our drawings, edge crossings can occur when long subtrees encroach on neighboring containment circles. An alternative method of allocating containment arcs might make it possible to  guarantee planar drawings.

Second, with our approach, remote descendants of the root can become vanishingly small on the viewing plane. Our system does give users a natural solution to this problem: selecting a different root node so as to allocate more space to its descendants~\cite{sarkar92,schaffer96}. However, future research could explore the algorithmic relation between our solution versus the distortion of the viewing plane in hyperbolic visualizations~\cite{lamping95,munzner95}. There are clearly differences: we position and move siblings by constraining them to circles on a parent-centered Euclidean plane, whereas hyperbolic layout algorithms position and move siblings through a non-Euclidean space. The relative computational and psychological merit of these different approaches, however, remains to be determined.

\begin{figure*}[t]
	\centering
	\small
	\subfigure[]{
		\label{fig:screenshot2.1}
		\includegraphics[width=.46\textwidth]{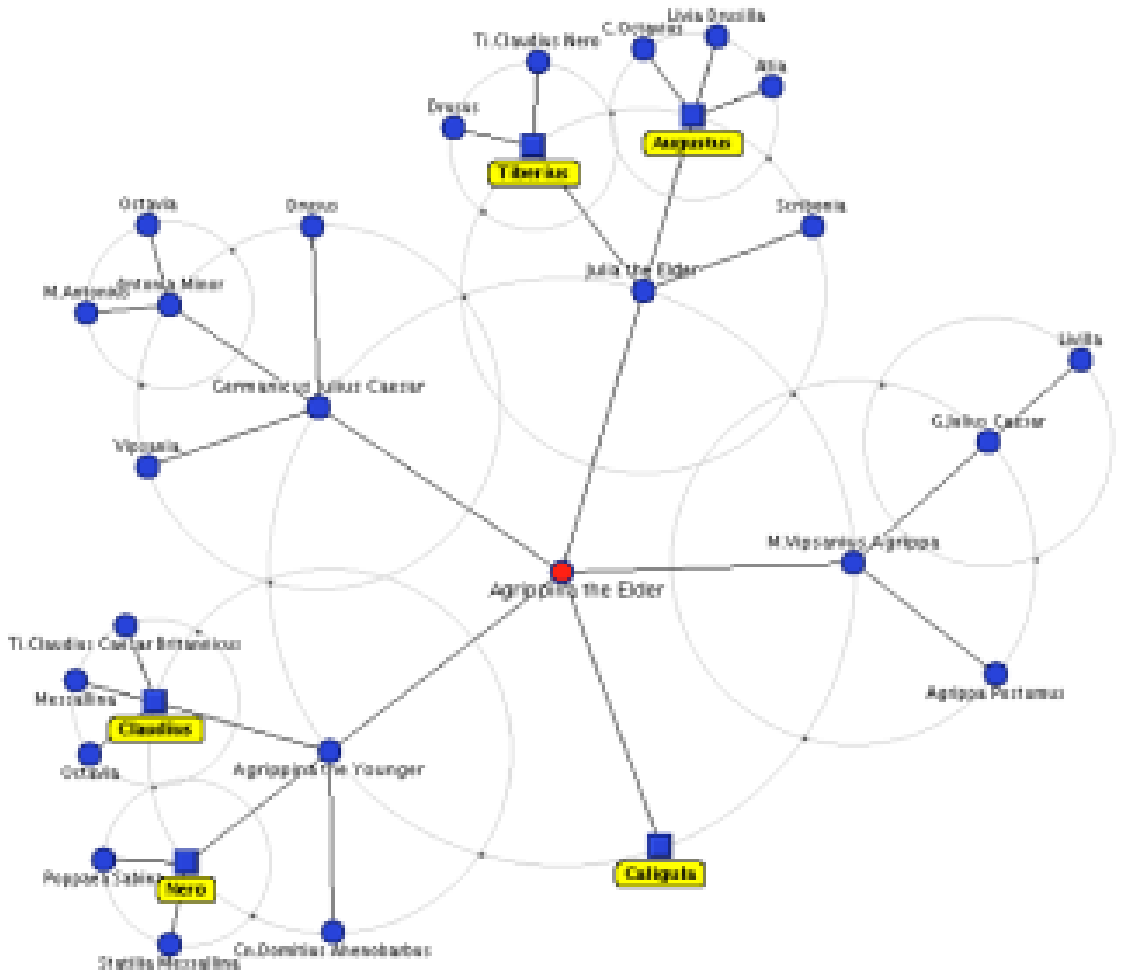}
	}
	\subfigure[]{
		\label{fig:screenshot2.2}
		\includegraphics[width=.46\textwidth]{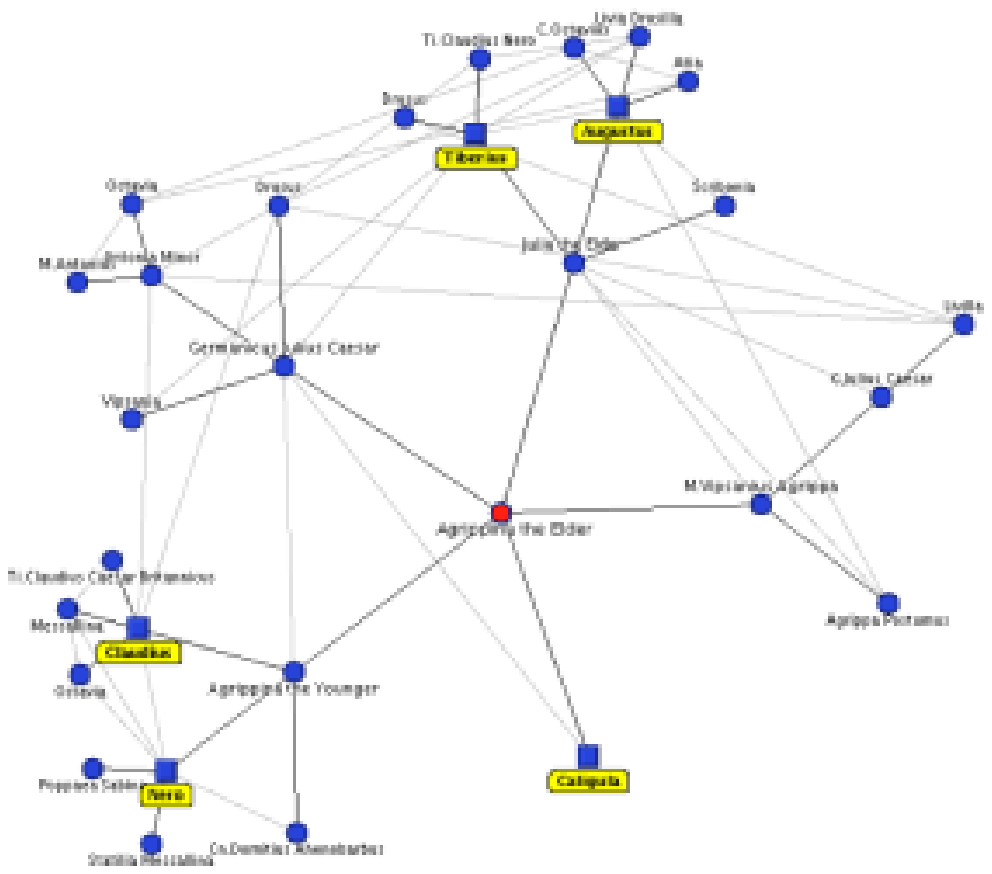}
	}
	\caption{An example drawing of the Julio-Claudian Imperial family network generated by our layout algorithm. An edge between two entities in the graph denotes either a marital, parental, or adoption relationship. The Roman emperors in the graph are indicated by boxed labels. The graph is shown in Figure~\ref{fig:screenshot2.2} with all the edges not included in the spanning tree in Figure~\ref{fig:screenshot2.1} revealed.}
	\label{fig:screenshot2}
\end{figure*}

\begin{figure*}[!t]
	\centering
	\small
	\subfigure[]{
		\label{fig:screenshot6.1}
		\includegraphics[width=.32\textwidth]{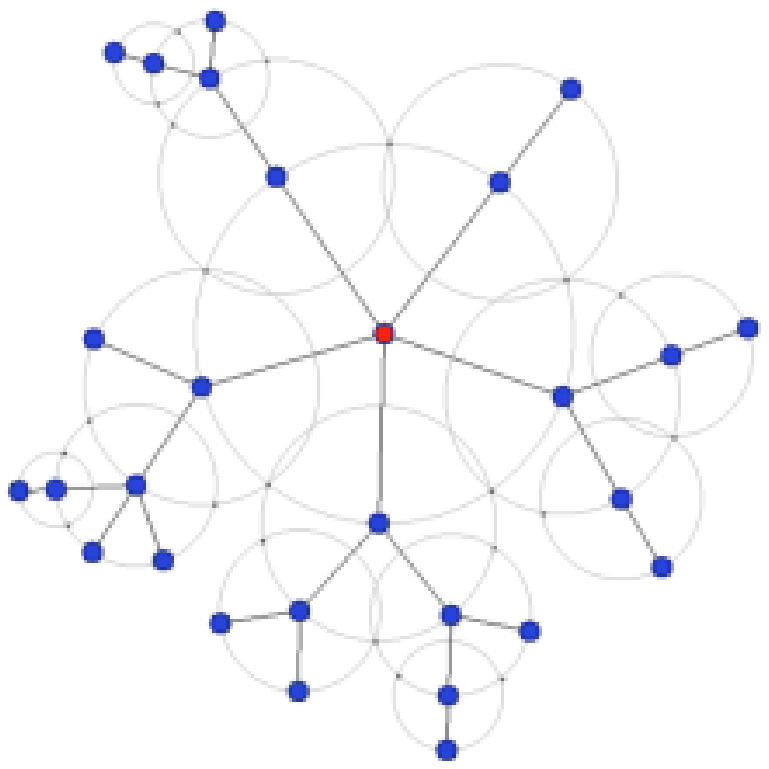}
	}
	\color{dividerColor}
	\line(0,1){150}
	\normalcolor
	\subfigure[]{
		\label{fig:screenshot6.2}
		\includegraphics[width=.32\textwidth]{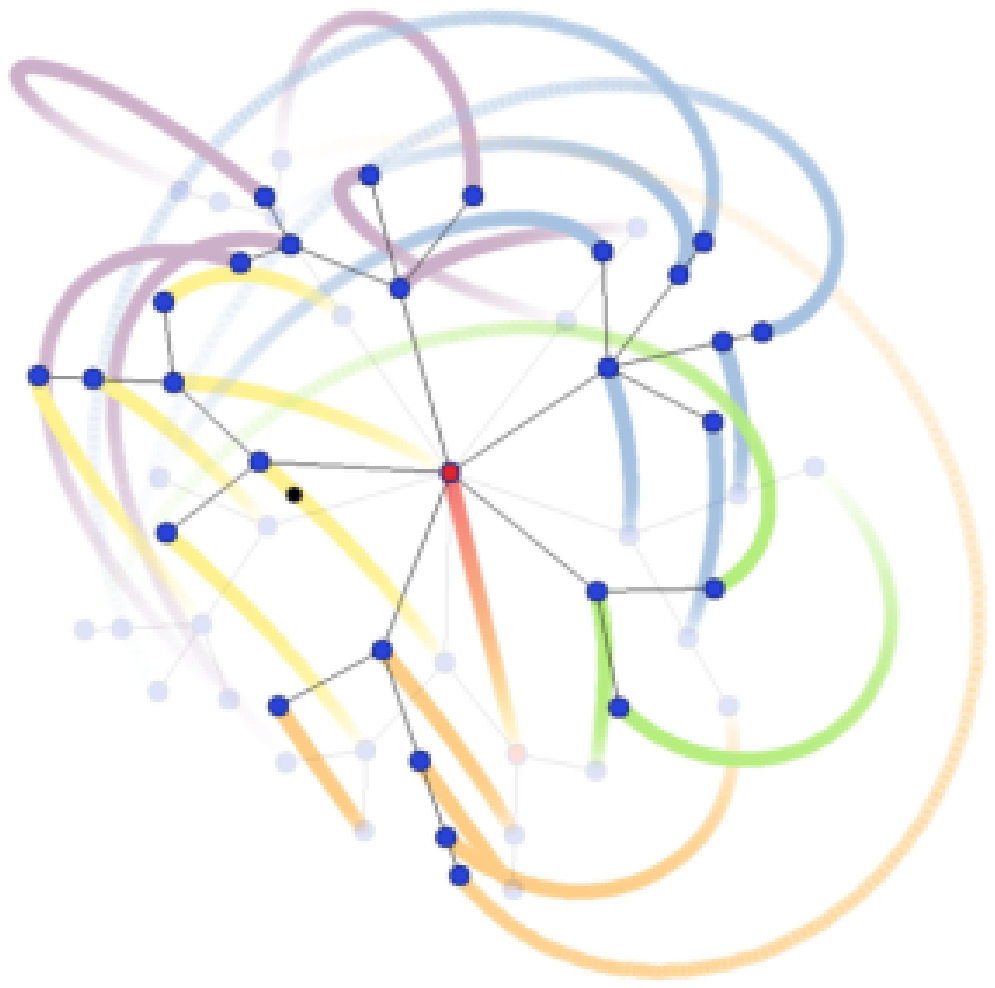}
	}
	\color{dividerColor}
	\line(0,1){150}
	\normalcolor
	\subfigure[]{
		\label{fig:screenshot6.3}
		\includegraphics[width=.32\textwidth]{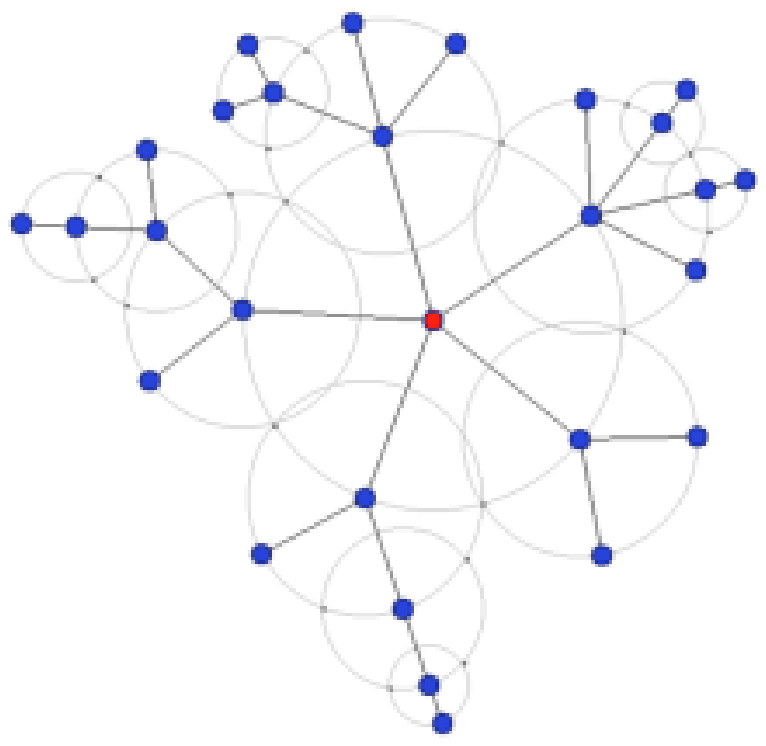}
	}
	\caption{An example animation sequence generated by our animation algorithm. The user selects a new root node from an existing graph drawing in Figure~\ref{fig:screenshot6.1} and the system transitions to the new layout in Figure~\ref{fig:screenshot6.3}. The trajectories of the nodes in the animation are highlighted in Figure~\ref{fig:screenshot6.2}.}
	\label{fig:screenshot6}
\end{figure*}

\section{Conclusion}
We have presented a radial graph layout visualization scheme based on a parent-centered data model for spanning trees extracted from a graph.  We introduced a static layout algorithm that produces drawings of graphs where the root's children are evenly spaced on a circle centered at the root and the children of nonroot nodes are evenly spaced on a semicircle emanating from their parent. We also introduced an animation algorithm that smoothly transitions a graph from one spanning-tree-based layout to another. We conducted experiments to compare our experimental system with Yee et al.'s  graph visualization system \cite{yee01}. The results from these experiments suggest that our visualization and animation schemes could indeed help users understand and explore graphs.

\bibliographystyle{plain}
\nocite{*}
\bibliography{infovis}
\end{document}